\documentclass[aps,pre,twocolumn,groupedaddress,showpacs,superscriptaddress,amssymb,amsmath]{revtex4-1}
\usepackage{graphicx}
\usepackage{tabularx}
\usepackage{color}
\usepackage{amsmath}
\usepackage{comment}
\newcommand{\be}{\begin{equation}}
\newcommand{\ee}{\end{equation}}
\newcommand{\bea}{\begin{eqnarray}}
\newcommand{\eea}{\end{eqnarray}}
\usepackage{dcolumn}
\usepackage{hyperref}
\usepackage{bm}
\usepackage{epsf}
\usepackage{subfigure}
\usepackage{epstopdf}%
\usepackage{amsfonts}

\let\la\langle \let\ra\rangle

\begin{document}

\title{Random walks on modular chains: Detecting structure through statistics}

\author{Matthew Gerry}
\affiliation{Department of Physics, University of Toronto, 60 Saint George St., Toronto, Ontario M5S 1A7, Canada}

\author{Dvira Segal}
\affiliation{Department of Physics, University of Toronto, 60 Saint George St., Toronto, Ontario M5S 1A7, Canada}
\affiliation{Chemical Physics Theory Group, Department of Chemistry and Centre for Quantum Information and Quantum Control,
University of Toronto, 80 Saint George St., Toronto, Ontario M5S 3H6, Canada}
\email{dvira.segal@utoronto.ca}
\date{\today}

\begin{abstract}
We study kinetic transport through 
modular networks consisting of alternating domains using both analytical and numerical methods.
We demonstrate that the mean velocity is {\it insensitive} to the local structure of the network,
and it indicates only on global, structural-averaged properties.
However, by examining high-order cumulants characterizing the kinetics, we reveal information on the degree of inhomogeneity of blocks and the size of repeating units in the network.  
Specifically, in  {\it unbiased} diffusion, the kurtosis is the first transport coefficient that 
exposes structural information, whereas in 
{\it biased} chains, the diffusion coefficient already reveals structural motifs.
Nevertheless, this latter dependence is weak and it disappears at both low and high biasing.
Our study demonstrates that high order moments of the population distribution over sites 
provide information about the network structure that is not captured by the first moment (mean velocity) alone.
These results are useful towards deciphering mechanisms and determining architectures
underlying long-range charge transport in biomolecules and biological and chemical reaction networks.
\end{abstract}
\maketitle

\section{Introduction}


Rolf Landauer's famous saying that ``the noise is the signal" \cite{RL} was originally stated in the context of noise measurements in electronic conductors, which can be used to reveal underlying many-electron interactions in the device. 
However, whether concerned  with electronic, chemical or biological kinetics, this statement captures a profound observation: that fluctuations of a signal can be used to probe the system---and more fundamentally so 
than the mean signal alone \cite{vankampen-book}.
Considering, for example, charge transport in conductors, the probably distribution function (PDF) 
to transfer $n$ electrons within a certain time encapsulates the full information 
on the charge current and its moments \cite{gernot-book,Kewming}. 
Similarly, the  PDF for first passage processes, e.g., in the context 
of protein folding or transport through ion channels,
contains rich information beyond what is conveyed by the averaged measure of the mean first passage time \cite{Zilman}. %

Chemical and biological reactions can be coarse-grained and modeled as network systems, and general principles can be understood by studying such systems in the context of stochastic thermodynamics \cite{udoCRN, rao2016, rao2018, avanzini2022}. The structure and topology of these networks can, for instance, be detected from the PDF of first-passage processes \cite{Kolom13, Kolom14,Kolom21,Kolom20}.
In particular, in chemical kinetics, details on reaction mechanisms can be inferred from the statistics of the kinetics. It was shown in Ref. \citenum{Busta13} that fluctuations in the time to complete an enzymatic cycle can be used to bound the number of intermediate steps. 
Recent studies continued and interrogated higher order moments of the PDF of the cycle completion time, beyond the mean and the variance. They found that
the skewness and kurtosis of this PDF reveal, e.g., whether the enzymatic network was unicyclic or multicyclic--information  that was not reflected  in the second moment of the PDF \cite{udo15,Barato22}. Investigations into the statistics of observable currents at steady state in systems described by kinetic networks have also revealed bounds on the relative fluctuations of the current determined explicitly by network structure \cite{pietzonka2016a,pietzonka2016b}. 


\begin{figure*} [ht]
\centering
\includegraphics[width=0.8\textwidth, trim = 30 30 30 30]{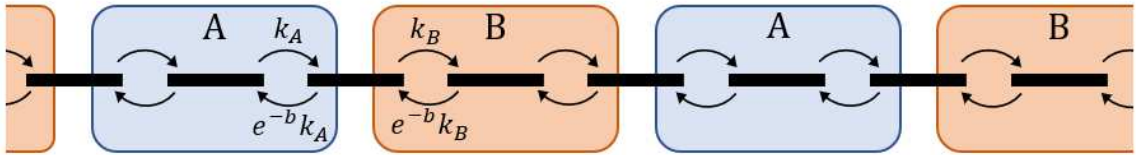}
\caption{Diagram of a modular chain on which an infinite random walk plays out. 
The unions of segments labeled ``A" and ``B" make up regions $A$ and $B$, respectively, 
where forward and reverse rates are given by 
Eqs.~(\ref{eq:fwd_rates}) and (\ref{eq:rev_rates}). 
For the example shown here the number of sites in each segment is $m_A=m_B=2$.}
\label{fig:chain_diagram}
\end{figure*}

In a different context, shot noise measurements of mesoscale, nanoscale, and atomic conductors provide
information on quantum transport that cannot be resolved from the electrical conductance itself \cite{Shot0}.
As such, shot noise experiments were used to characterize, e.g., the fractional quantum Hall effect \cite{Shot1}, electron transport in the Kondo regime \cite{Shot2}, electron-phonon interaction effects in
molecular junctions \cite{Shot3,Shot4}, and structural and energetic asymmetry of atomic-scale conductors \cite{Shot5}.
Beyond shot noise, which corresponds to the second moment of carriers flow, it was recently shown that the skewness and kurtosis of the PDF of charge transport reveal information on the violation of the thermodynamic uncertainty relation (TUR) \cite{bijay1,bijay2} and the
impact of many-body effects in transport \cite{Ptaz22}.

The objective of this study is to identify what new, detailed information on the structure of a kinetic
network can be revealed from successive high moments of the 
distribution of population, beyond its mean.  
We focus on random walk in one-dimensional networks, which are either homogeneous or modular.
An example of a modular system is depicted in Fig. \ref{fig:chain_diagram}; the configurational space map of the model is presented in Fig. \ref{fig:cycle_diagram}.
%
It was previously shown, through numerical simulations of quantum dissipative transport, 
that the average electron current through modular junctions 
only depends on the {\it average} resistance of the chain, and it is insensitive to the local structure \cite{Francisco}.
These results were demonstrated numerically using two approaches, B\"uttiker probes' simulations \cite{Roman} and quantum rate equations of the Lindblad form.
One of the goal of the present study is to derive an analogous result analytically, which we do using classical kinetics for modular networks with unit block size.
More generally, our aim here is to reveal with analytical and numerical work information about the local structure of a network (e.g., inhomogeneity of blocks and size of repeating units)  from cumulants of the distribution (mean, diffusion coefficient, skewness, and kurtosis). 

We discover that in an {\it unbiased} random walk on a modular network, the kurtosis is the {\it first}
cumulant 
that expresses local structural information.
In contrast, in {\it biased} random walks, the diffusion coefficient can reveal structural information, but this dependence is rather weak. The sensitivity to the local structure amplifies only in the skewness and more so in the kurtosis.

The paper is organized as follows. We present the 
random walk models, in real space and configuration space in Sec. \ref{sec:MM}.
We study the  statistics of carriers transport in Sec. \ref{sec:fcs}. Analytic results for the first four cumulants of the population distribution  
for modular chains with a block size one 
are presented in Sec. \ref{sec:analytic}. Complementing this analysis, numerical simulations 
are included in Sec. \ref{sec:sim}. 
Real-space simulations of the random walk are presented in Sec. \ref{sec:space}, revealing the rich structure of the PDF in modular models.
We discuss our results and conclude in Sec. \ref{sec:summ}.

\section{Models}
\label{sec:MM}

\begin{figure}[]
 \centering
 \includegraphics[width=0.8\columnwidth, trim=30 20 30 20]{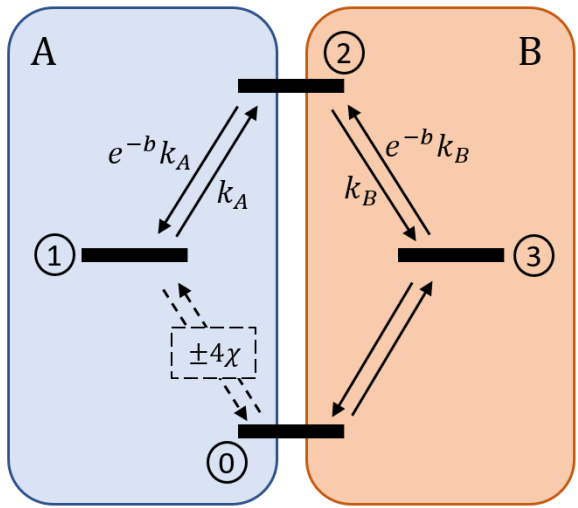}
\caption{The modular chain of Fig. \ref{fig:chain_diagram} can be mapped into a
a finite cyclic network of states.
This cyclic network is created by recognizing the periodicity in the structure of the modular infinite chain. The clockwise direction around the cycle corresponds to the forward direction of the random walk. 
Full counting statistics tracks the net number of times the system completes the $0\rightarrow1$ transition, 
as indicated by the dashed arrows and counting field. 
This number is then scaled up by $m_P=4$ to count the number of steps ``away" from the initial site at steady state.}
\label{fig:cycle_diagram}
\end{figure}

We study random walks in structured networks corresponding to two setups: (I) An infinite one-dimensional modular lattice with repeating segments, Fig. \ref{fig:chain_diagram}. (II) A finite unicyclic network, Fig. \ref{fig:cycle_diagram}, representing reaction kinetics. Alternatively, the cyclic model corresponds to the configurational space of the infinite chain model (I). These models help to highlight the effects of underlying periodic structure, which has been investigated in numerous past works for random walks as well as diffusion in continuous space \cite{dieterich77,Derrida, dicrescenzo,Reimann01,Reimann08,tracerD,Illien}.

\subsection{Modular and homogeneous chains}
\label{sec:chains}

We consider a continuous-time random walk on an infinite modular chain, see Fig. \ref{fig:chain_diagram}. In our model,
transition rates alternate between different values for segments some number of sites long. 
Most generally, segments associated with the two sets of transition rates are $m_A$ and $m_B$ sites long, 
where $m_A$ and $m_B$ may or may not be equal. 
The total period for this alternation of transition rates is then 
$m_P=m_A+m_B$. 
The blocks are made distinct by using different hoping rates between sites within segment A and within segment B.

This network can represent a polymer with alternating units, stiff and flexible. Such modular structures may be used to model long-range charge transport in polymers and biomolecules \cite{Geo1,Geo2,Beratan19}. The object of our calculation is the PDF to find the system $n$ sites away from its starting point, within some time. We refer to the first four cumulants of this PDF, scaled by time, as the mean velocity, diffusion coefficient, skewness, and kurtosis.


The forward transition rate from site $i$ to site $i+1$ is given by
\bea\label{eq:fwd_rates}
k_{i+1,i} = 
\begin{cases}
k_A\equiv \frac{\tau^2}{\gamma_A},\:\:  (i\ \textrm{mod } m_P) <m_A \\
k_B\equiv \frac{\tau^2}{\gamma_B},\:\: (i\ \textrm{mod } m_P) \geq m_A.
\end{cases}
\eea
Here, we apply the modulo (mod) operation. 
We note here that we start the site-counting at zero, which is an A-type site, and that there is a site for every $i\in \mathbb{Z}$. 
$\tau$ is a constant which gives rise to an overall scale for the rates. 
The rates differ due to the difference in the values of $\gamma_A$ and $\gamma_B$, 
which 
serve a function analogous to resistance.

Note that in the corresponding quantum transport model \cite{Francisco}, 
the blocks A and B were made distinct by setting different local decoherence rates for electrons,  
being  high or low relative to the tunneling energies $\tau$. A segment with high (low) decoherence corresponds to a flexible (rigid) region in a polymer \cite{Segal00,Cao}.

Assuming local detailed balance, 
given the forward transition rates, the reverse rates are given by \be\label{eq:rev_rates}
    k_{i,i+1} = e^{-b}k_{i+1,i},
\ee
where $b$ represents a uniform bias. 
In the context of a charged particle hopping under an external electric field, the bias is $b=\Delta \mu/(k_bT)$, with $\Delta \mu$ the local potential bias, $T$ the temperature, and $k_B$ as the Boltzmann constant.
%
Each individual transition is symmetric in the case that $b=0$, and the walk is biased in the forward direction for $b>0$. In what follows we assume the bias is equal between every two sites. 

To contrast the  modular chain, we build a corresponding homogeneous chain 
whose forward transition rate at all sites is given by $k^* = \tau^2/\gamma^*$, 
with $\gamma^*\equiv(m_A\gamma_A + m_B\gamma_B)/m_P$ (the average of $\gamma_A$ and $\gamma_B$ weighted by their 
respective segment lengths). 
The reverse rate is uniformly given by this rate scaled by $e^{-b}$, in analogy to the above.

We will focus primarily on the case that the segment lengths are equal, $m_A=m_B\equiv m$ such that $m_P=2m$. Accordingly, $k^*$ is calculated simply using the arithmetic mean value, $\gamma^* = \Bar{\gamma} = (\gamma_A + \gamma_B)/2$; $k^*=\tau^2/\bar\gamma$.

\subsection{Bipartite finite cycle}


In addition to the infinite modular chain and its associated homogeneous counterpart, we consider a closely related {\it finite} network consisting of a single cycle with $m_P$ sites, see Fig. \ref{fig:cycle_diagram}.
This cyclic kinetic network can be viewed as a reaction network with a total of $m_P$ internal states: The first $m_A$ steps are slow, and they are followed by $m_B$ fast steps.
At the end of each kinetic cycle, a product specie is generated, and thus the completion of a cycle can be monitored. The probability distribution for the diffusion process along the chain now describes the statistics of the number of product molecules formed. 

Furthermore, the finite cycle can be constructed from the infinite chain discussed in Sec. \ref{sec:chains}
by connecting the right-most site of a `B' segment back to the left-most site of the previous `A' segment. 
The transition rates on this cyclic network can be defined in accord
using Eqs.~(\ref{eq:fwd_rates}) and (\ref{eq:rev_rates}), but replacing each index $i$ with  ($i$ mod $m_P$),
such that neither ever exceeds $m_P-1$, but instead cycles back to 0. 
For example, if $m=2$ ($m_P=4$), the  unicyclic network we associate with the random walk consists of four states (labeled $i=0,1,2,3$, as shown in Fig. \ref{fig:cycle_diagram}), with $k_{1,0}=k_{2,1}=k_A$, $k_{3,2} = k_{0,3} = k_B$, and reverse rates equal to these forward rates scaled by $e^{-b}$. $k_{i,j}=0$ for any $|i-j|\geq2$, with the exception of the last-to-first transition closing the loop. 
For comparison, we also study the associated homogeneous cycle by setting all forward rates to $k^*$.

We build the random walk on the cyclic network using  exactly the same transition rates as one on the infinite modular chain. As a result, the statistics of the net number of steps taken in the clockwise direction follows the exact same statistics as the net number of steps taken 
on the corresponding infinite chain. We use the variable $n$ to represent both of these quantities. 


The main objective  of this work is to investigate how signatures of modular structure, 
in comparison to the homogeneous structure, manifest in the statistics of random walks on these chains, at steady state.
Useful information on the structure that we would like to reveal is (i) the degree of inhomogeneity, $\Delta \gamma= \gamma_A-\gamma_B$,
and (ii) the block size. 


\section{Full counting statistics for forward steps}
\label{sec:fcs}


We are interested in the scaled cumulants for the distribution over $n$ in the long-time limit,
\be\label{eq:cumulants_def}
    \mathcal{C}_k\equiv \lim_{t\rightarrow\infty}\frac{\la\la n^k(t)\ra\ra}{t},
\ee
where $n(t)$ is the stochastic variable representing the site of the walker at time $t$, such that $\langle\langle n^k(t)\rangle\rangle$ is the $k^{th}$ cumulant of the distribution over this quantity. We may suppose $n(0)=0$, though for large enough $t$ the initial conditions have no impact on the scaled cumulants.

Our implementation of full counting statistics is based on the equivalence between the statistics of $n$ for both the infinite modular chain and the finite cycle introduced in Sec. \ref{sec:MM}. This equivalence arises from the periodicity of the modular structure.
Whether we are ultimately interested in the infinite chain or the finite cycle, we carry out full counting statistics by writing down the $m_P\times m_P$ rate matrix, $\mathcal{L}$, for the $m_P$-state unicyclic network. The off-diagonal elements $\mathcal{L}_{ij}$ are the transition rates $k_{i,j}$ and the diagonal elements are set such that the columns sum to zero \cite{vankampen-book}. 

Knowledge of the matrix elements of $\mathcal{L}$ allows us write the rate of change of the probability $\mathcal{P}_i(\mathcal{N};t)$ for the system to be found at site $i$, given that it has completed exactly $\mathcal{N}$ rounds 
around the cycle. We choose to register trips around the cycle at the $0\leftrightarrow1$ transition, such that \cite{qar2018}
\begin{align}
    \frac{d}{dt}\mathcal{P}_0(\mathcal{N};t) = &-\mathcal{P}_0(\mathcal{N};t)(k_{1,0} + k_{m_P-1,0})
    \nonumber\\
    &+ \mathcal{P}_{m_P-1}(\mathcal{N};t)k_{0,m_P-1}
    \nonumber\\
    &+ \mathcal{P}_1(\mathcal{N}+1;t)k_{0,1},
    \nonumber\\
    \frac{d}{dt}\mathcal{P}_1(\mathcal{N};t) = &-\mathcal{P}_1(\mathcal{N};t)(k_{0,1} + k_{2,1})
    \nonumber\\
    &+ \mathcal{P}_0(\mathcal{N}-1;t)k_{1,0}
    \nonumber\\
    &+ \mathcal{P}_2(\mathcal{N};t)k_{1,2},
    \nonumber\\
    \frac{d}{dt}\mathcal{P}_j(\mathcal{N};t) = &-\mathcal{P}_j(\mathcal{N};t)(k_{j\oplus 1,i} + k_{j\ominus 1,j})
    \nonumber\\
    &+ \mathcal{P}_{j\ominus1}(\mathcal{N};t)k_{j,j\ominus1}
    \nonumber\\
    &+ \mathcal{P}_{j\oplus1}(\mathcal{N};t)k_{j,j\oplus1},
\end{align}
where the last equation applies to the case where $j\neq 0,1$, and $\oplus$ and $\ominus$ represent addition and subtraction modulo $m_P$, respectively.

We then introduce a counting field $\chi$ and Fourier transform these quantities with respect to $\mathcal{N}$, leading to the $m_P$-element 
characteristic function $\mathcal{Z}(\chi;t)$, whose elements are given by
\be
    \mathcal{Z}_j(\chi;t) = \sum_{\mathcal{N}=-\infty}^\infty\mathcal{P}_j(\mathcal{N};t)e^{im_P\chi}.
\ee    
We include a factor of $m_P$ in the exponent so that the counting effectively tracks the number, $n$, of forward steps, rather than the number, $\mathcal{N}$, of complete cycles. At sufficiently long times, $n\approx m_P\mathcal{N}$. Getting a $\chi$-dressed (or ``tilted") rate matrix, $\mathcal{L}(\chi)$, to time-evolve $\mathcal{Z}(\chi;t)$ directly, amounts to multiplying the transition rate $k_{1,0}$ as it appears in $\mathcal{L}$ by a factor $\exp(im_P\chi)$, and multiplying $k_{0,1}$ by the reciprocal, $\exp(-im_P\chi)$\cite{gernot-book}. Note that an alternative approach towards obtaining $\mathcal{L}(\chi)$ would be to multiply \textit{every} clockwise transition rate in $\mathcal{L}$ by a factor of just $\exp(i\chi)$ and \textit{every} anticlockwise transition rate by a factor of $\exp(-i\chi)$; for full counting statistics of $n$ in the transient regime this would give greater detail, but it is equivalent at steady state.

The scaled cumulant generating function is given by \cite{qar2018}
\be 
    \mathcal{G}(\chi) = \lim_{t\rightarrow\infty}\frac{1}{t}\ln\sum_j\mathcal{Z}_j(\chi;t).
\ee
This is equivalent to the ``dominant" eigenvalue of $\mathcal{L}(\chi)$: that whose real part approaches zero in the limit that $\chi$ approaches zero. This function may be used to derive all of the scaled cumulants at steady state via the relation 
\cite{gernot-book},
\be 
    \mathcal{C}_k = \frac{d^k}{d(i\chi)^k}\mathcal{G}(\chi)\bigg|_{\chi=0}.
\ee
As such, the method outlined above is sufficient to fully characterize the steady-state probability distribution for walkers 
through systems modeled by the type of modular random walk we have introduced.

\subsection{Analytic results for $m_P=2$}
\label{sec:analytic}

We  focus here on the special case of the modular chain where $m=1$ (i.e. all segments are one site long and the total period is size $m_P=2$). The differing rates for the two regions are based on differing values of $\gamma_A$ and $\gamma_B$. 
Without any significant loss of generality, we may suppose $\gamma_A\geq\gamma_B$. We want to investigate the statistics of this random walk at steady state, with particular focus on how the values of the scaled cumulants depend on $\Delta\gamma=\gamma_A-\gamma_B$. The limit $\Delta\gamma\rightarrow0$ represents the case where the modular chain becomes identical to its homogeneous counterpart.

The counting field-dependent rate matrix for the cyclic network associated with this modular random walk is
\be
    \mathcal{L}(\chi) = 
    \left(
    \begin{array}{cc}
        -k_A-e^{-b}k_B & e^{-b-2i\chi}k_A + k_B \\
        e^{2i\chi}k_A + e^{-b}k_B & -e^{-b}k_A - k_B
    \end{array}
    \right).
\ee
Note that setting $m=1$ gives rise to the special case where two consecutive forward steps of the random walk correspond to two transitions between the {\it same} pair of states in the finite network. Accordingly, the elements of $\mathcal{L}$ contain sums of two transition rates.

Following the method outlined in Sec. \ref{sec:fcs}, we obtain the scaled CGF,
\begin{widetext}
\be\label{eq:cgf}
    \mathcal{G}(\chi) = 2\frac{\tau^2}{\Bar{\gamma}}e^{-\frac{b}{2}}\frac{\sqrt{\cosh^2(\frac{b}{2} + i\chi) -\frac{1}{4}(\frac{\Delta\gamma}{\Bar{\gamma}})^2\sinh(i\chi)\sinh(b+i\chi)}-\cosh(\frac{b}{2})}{1 - \frac{1}{4}(\frac{\Delta\gamma}{\Bar{\gamma}})^2}.
\ee
\end{widetext}
We remind the reader that the random walk is defined
in terms of the rates $k_A=\frac{\tau^2}{\gamma_A}$ and $k_B=\frac{\tau^2}{\gamma_B}$ with the structural asymmetry
$\Delta\gamma=\gamma_A-\gamma_B$, further defining the averaged measure for resistance $\bar \gamma=\frac{\gamma_A+\gamma_B}{2}$,
and the corresponding rate constant, $k^*=\frac{\tau^2}{\bar\gamma}$.

\subsubsection{Mean velocity}
Taking the first derivative of $\mathcal{G}(\chi)$ and setting $\chi=0$, we obtain an expression for the first cumulant, the mean velocity, at steady state,
\be\label{eq:mean}
    \mathcal{C}_1 = 2\frac{\tau^2}{\Bar{\gamma}}e^{-\frac{b}{2}}\sinh\bigg(\frac{b}{2}\bigg) = 2e^{-\frac{b}{2}}\sinh\bigg(\frac{b}{2}\bigg)k^*.
\ee
Upon inspection, we note immediately that this quantity is completely independent of the difference $\Delta\gamma$, and is therefore the same for a random walk on the modular chain as it is for the homogeneous chain with the appropriately defined rate $k^*$. In fact, using the results of Ref. \cite{Derrida}, this form of the mean velocity can be derived for the modular random walks considered in this work, independent of the spatial period $m_P$, as long as $m_A=m_B$. We stress that an experimental investigation of a system modeled by such a random walk {\it cannot detect the underlying modular structure in any way if it takes into consideration only the first cumulant}.

Checking the limits of high and low $b$, we see expected behavior. Namely, for small bias $b$, a linear response behavior shows, $\mathcal{C}_1\rightarrow k^*b$, which goes strictly to zero when $b=0$. For large $b$, reverse transitions are suppressed and $\mathcal{C}_1\rightarrow k^*$, which is the forward rate for the homogeneous chain. These trends are displayed in Fig. \ref{fig:mean} for modular chains of varying periods $m_P\geq2$ based on numerical simulations. 

\subsubsection{Diffusion coefficient}

Following the same procedure but taking the second order derivative, we get an expression for the second cumulant, or the diffusion coefficient,
\begin{align}\label{eq:variance}
    \mathcal{C}_2 = 2\frac{\tau^2}{\Bar{\gamma}}\frac{e^{-\frac{b}{2}}}{\cosh(\frac{b}{2})}\bigg[&\cosh^2\bigg(\frac{b}{2}\bigg)
    \nonumber\\
    &\:\:+ \frac{1}{4}\sinh^2\bigg(\frac{b}{2}\bigg)\bigg(\frac{\Delta\gamma}{\Bar{\gamma}}\bigg)^2\bigg]. 
\end{align}
Unlike the mean, this quantity depends on $\Delta\gamma$, and could be used as a probe of modular structure. We note, in particular, that a modular chain always exhibits a {\it greater} diffusion coefficient than its homogeneous counterpart. However, this dependence appears in a term proportional to $e^{-b/2}\sinh(b/2)\tanh(b/2)$, thus its impact is most prominent for heavily biased walks and diminishes in the limit $b\rightarrow0$. Expanding the full expression in powers of $b$, we would see $\Delta\gamma$-dependence only in terms of order-$b^2$ and higher. For an unbiased random walk, $b=0$, this dependence disappears completely, and we have the familiar result that $\mathcal{C}_2=2k^*$. This may be compared to the large $b$ limit, where
\be 
    \mathcal{C}_2 \xrightarrow{b\to\infty} \bigg[1 + \frac{1}{4}\bigg(\frac{\Delta\gamma}{\Bar{\gamma}}\bigg)^2\bigg]k^*.
\ee
Here, varying $\Delta\gamma$ between zero and its maximum possible value of $2\Bar{\gamma}$ can lead $\mathcal{C}_2$ to vary by a factor of 2.

To probe the behavior of $\mathcal{C}_2$ with varying bias in greater detail, we return to the exact expression given in Eq.~(\ref{eq:variance}) and note that at low $b$, $\frac{\partial \mathcal{C}_2}{\partial b}<0$. 
However, on the condition that $\Delta\gamma/\Bar{\gamma}>2/\sqrt{3}$, there exists a value of $b$ at which this derivative becomes positive. This leads $\mathcal{C}_2$ to increase with increasing $b$ at sufficiently high values of $b$. This behaviour, which is counter to the conventional understanding of the effect of bias, could serve as experimentally accessible evidence of underlying modular structure.

Our analytical results for $m_P=2$ reveal that (i) the diffusion coefficient monotnonically grows with $\Delta \gamma$ and (ii)
for modular systems, it may display a nonmonotonic behavior with bias. 
These trends, as well as the enhancement of the diffusion coefficient with the lattice period $m_P$ are presented in Fig. \ref{fig:var} using numerical simulations. 

\subsubsection{Skewness and kurtosis}
Taking the third-order derivative of the CGF with respect to $\chi$ we obtain the skewness, which is given by
    

\begin{align}\label{eq:skewness}
    \mathcal{C}_3 = 2\frac{\tau^2}{\Bar{\gamma}}\frac{e^{-\frac{b}{2}}\sinh(\frac{b}{2})}{\cosh^2(\frac{b}{2})}\bigg[ &\cosh^2\bigg(\frac{b}{2}\bigg)+ \frac{3}{4}\bigg(\frac{\Delta\gamma}{\Bar{\gamma}}\bigg)^2 
    \nonumber\\
    &+ \frac{3}{16}\sinh^2\bigg(\frac{b}{2}\bigg)\bigg(\frac{\Delta\gamma}{\Bar{\gamma}}\bigg)^4 \bigg].
\end{align}

Like the first cumulant, this quantity vanishes completely in the absence of bias. However, $\Delta\gamma$-dependence is present in contributions that are first-order in $b$. Therefore, at low but finite bias (Fig. \ref{fig:skewness}(a)), 
the skewness may be said to express the modular structure of the chain more substantially than the diffusion coefficient does.

In the opposite limit that $b\rightarrow\infty$, the skewness saturates to the finite value,
\be 
    \mathcal{C}_3 \xrightarrow {b\to\infty} \bigg[1 + \frac{3}{16}\bigg(\frac{\Delta\gamma}{\Bar{\gamma}}\bigg)^4 \bigg]k^*,
\ee
see  Fig. \ref{fig:skewness}(b). 
This value exhibits fourth-order dependence on $\Delta\gamma$, in comparison to the second-order dependence of the diffusion coefficient on $\Delta\gamma$ in this regime. For very small $\Delta\gamma/\Bar{\gamma}$, we expect the skewness to be less sensitive to the modular structure than the diffusion coefficient is.

Therefore, whether the diffusion coefficient or skewness better expresses underlying modular structure is dependent on the situation--if bias is high and the properties of the two regions are believed to differ only slightly, measurements of $\mathcal{C}_2$ are likely the more expressive option. If the bias is weak, the skewness is more sensitive to changes in $\Delta\gamma$ and will likely convey more about the structure.
The  skewness in the present model is positive; it is interesting to devise related models that display a negative skewness \cite{Salazar}. 

The kurtosis is given by the fourth-order derivative of $\mathcal{G}(\chi)$ evaluated at $\chi=0$,
\begin{widetext}
\be
    \mathcal{C}_4 = 2\frac{\tau^2}{\Bar{\gamma}}\frac{e^{-\frac{b}{2}}}{\cosh^3(\frac{b}{2})}\bigg[ \cosh^4\bigg(\frac{b}{2}\bigg) + f(b)\bigg(\frac{\Delta\gamma}{\Bar{\gamma}}\bigg)^2 - g(b)\bigg(\frac{\Delta\gamma}{\Bar{\gamma}}\bigg)^4 + \frac{15}{64}\sinh^4\bigg(\frac{b}{2}\bigg)\bigg(\frac{\Delta\gamma}{\Bar{\gamma}}\bigg)^6 \bigg],
\ee
\end{widetext}
where $f(b) = \frac{1}{64}(e^{-2b}-36e^{-b} + 118 - 36e^{b} + e^{2b})$ and $g(b) = \frac{9}{256}(e^{-2b}-12e^{-b}+22-12e^{b}+e^{2b})$. Continuing the pattern that has emerged, we see more complex dependence on $\Delta\gamma$, now up to order-$\Delta\gamma^6$.

Specifically, even at zero bias, the kurtosis exhibits $\Delta\gamma$-dependence, taking the form
\be 
    \mathcal{C}_4 \xrightarrow{b\to0} 2\bigg[ 1 + \frac{3}{4}\bigg(\frac{\Delta\gamma}{\Bar{\gamma}}\bigg)^2\bigg]k^*.
\ee
This makes the kurtosis the lowest-order cumulant to be nonvanishing and express information about the modular structure even in the completely unbiased case.

In the infinite bias limit, the kurtosis goes to
\be 
    \mathcal{C}_4\xrightarrow{b\to \infty} \bigg[1 + \frac{1}{4}\bigg(\frac{\Delta\gamma}{\Bar{\gamma}}\bigg)^2 - \frac{9}{16}\bigg(\frac{\Delta\gamma}{\Bar{\gamma}}\bigg)^4 + \frac{15}{64}\bigg(\frac{\Delta\gamma}{\Bar{\gamma}}\bigg)^6\bigg]k^*.
\ee
In sum, this analytical work and numerical simulations show that (i) the kurtosis grows with the degree of modularity $\Delta \gamma$ even at zero bias (Fig. \ref{fig:kurtosis}(a)), and  that (ii) unlike the homogenous case, in modular chains it can become negative, Fig. \ref{fig:kurtosis}(b)-(c).

We have shown that for the case where $m=1$, varying amounts of information about the probability distribution over the quantity $n(t)/t$ at steady state are needed to discern underlying modular structure of the chain.  
The first cumulant, or mean velocity, is always identical to that for the analogous homogeneous chain. Therefore, at the very least, measurements of the second cumulant are needed to establish whether or not the transition rate alternates between two values, and, if so, the degree to which the values $\gamma_A$ and $\gamma_B$ differ. However, depending on the specific features of the random walk in question (i.e. if the bias is too low, or $\Delta\gamma/\Bar{\gamma}$ is large), measurements of the skewness or kurtosis may be more effective.



\subsection{Ratios of cumulants and the TUR}
We note further that these expressions lend themselves to being studied in the context of their ratios. One such ratio is the Fano factor (relative fluctuations, or ``noise-to-signal" ratio). This quantity is known to be bounded in a manner dependent on the kinetic network structure \cite{pietzonka2016a}. Futhermore, it is significant to studies of the TUR, a cost-precision tradeoff which bounds it terms of the entropy production rate \cite{udoTUR,ging16,juan17}, as well as the kinetic uncertainty relation, which bounds it in relation to the dynamical activity
\cite{terlizzi2018,kewmingKUR,vo2022}.
Evaluating the Fano factor for the modular random walk with $m_P=2$, we note that prefactors on the expressions for the cumulants cancel out, leaving,
\be
    \frac{\mathcal{C}_2}{\mathcal{C}_1} = \coth\bigg(\frac{b}{2}\bigg) + \frac{1}{4}\tanh\bigg(\frac{b}{2}\bigg)\bigg(\frac{\Delta\gamma}{\Bar{\gamma}}\bigg)^2.
\ee

If we suppose the random walk to represent a physical system, with nonequilibrium conditions giving rise to the bias $b$, then it is consistent to suppose that the entropy production rate is given by $\la \sigma\ra = \mathcal{C}_1b$. We can get an exact expression for the TUR ratio, $\la \sigma\ra\mathcal{C}_2/\mathcal{C}_1^2$, which is required to be greater than or equal to 2,
\begin{align}
    \la \sigma\ra\frac{\mathcal{C}_2}{\mathcal{C}_1^2} &= b\frac{\mathcal{C}_2}{\mathcal{C}_1} = b\coth\bigg(\frac{b}{2}\bigg) + \frac{1}{4}b\tanh\bigg(\frac{b}{2}\bigg)\bigg(\frac{\Delta\gamma}{\Bar{\gamma}}\bigg)^2.
\end{align}
This expression does indeed take on values greater than or equal to 2 for all values of $b$, going to exactly 2 strictly in the limit $b\rightarrow0$. In addition, we note that the second term represents a nonnegative contribution present only in the case of a modular chain with nonzero bias. As such, a random walk on a modular chain never comes as close to saturating the TUR as its homogeneous counterpart in the presence of bias.

Turning to the higher-order cumulants, we note that the ratio $\mathcal{C}_4/\mathcal{C}_3$ exhibits the exact same behavior in the absence of modular structure. However, it exhibits a more complicate dependence on $\Delta\gamma$. For instance, if we focus on the regime where $\Delta\gamma$ is small compared to $\Bar{\gamma}$, $\mathcal{C}_4/\mathcal{C}_3$ is given to good approximation by its expansion to second order in $\Delta\gamma/\Bar{\gamma}$. Accordingly,
\begin{align} 
    \frac{\mathcal{C}_4}{\mathcal{C}_3}=&\coth\bigg(\frac{b}{2}\bigg)\bigg[1 + \frac{1}{\cosh^2(\frac{b}{2})}\bigg(\frac{f(b)}{\cosh(\frac{b}{2})}-\frac{3}{4}\bigg)\bigg(\frac{\Delta\gamma}{\Bar{\gamma}}\bigg)^2\bigg] 
    \nonumber\\
    &+ \mathcal{O}\bigg(\bigg(\frac{\Delta\gamma}{\Bar{\gamma}}\bigg)^4\bigg).
\end{align}
Unlike in the case of the relative fluctuations, the contribution from $\Delta\gamma/\Bar{\gamma}$ takes on negative values at low, finite $b$. This means that modularity may suppress the ratio $\mathcal{C}_4/\mathcal{C}_3$ below the values it would take on in the homogeneous case, and even below $2/b$, distinguishing the behavior of this quantity from that of the standard TUR ratio.

\begin{figure}
    \centering
\includegraphics[width=0.8\columnwidth, trim=15 10 15 15]{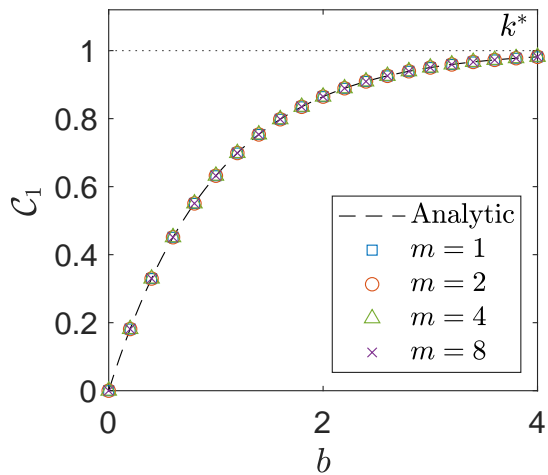}
\caption{The first cumulant of the random walk process, corresponding to the mean velocity,
presented as a function of the bias $b$ for varying segment lengths, with $m_A=m_B\equiv m$ such that $m_P=2m$. 
The dashed curve represents the expression given in Eq.~(\ref{eq:mean}). $\tau=\Bar{\gamma}=\Delta\gamma=1$.}
\label{fig:mean}
\end{figure}
\begin{figure}[htbp]
    \centering
\includegraphics[width=0.85\columnwidth, trim=18 18 18 18]{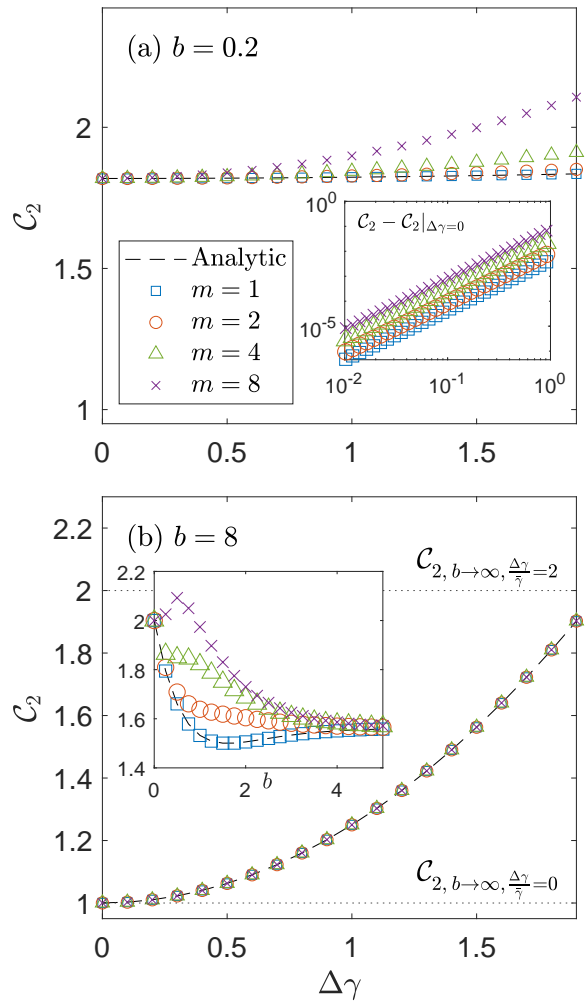}
\caption{Diffusion coefficient as a function of $\Delta\gamma$ for varying segment lengths, $m_A=m_B\equiv m$. Dashed black curves represent the analytic result for $m_P=2$. Plots are shown for (a) low and (b) high bias. The inset of panel (a) depicts the same quantity on a log-log plot, with the $\Delta\gamma=0$ contribution subtracted off, demonstrating that the diffusion coefficient scales quadratically with $\Delta\gamma$ for varying $m$. The inset of panel (b) shows $\mathcal{C}_2$ as a function of $b$ for $\Delta\gamma=1.5$, demonstrating its nonmonotinicity with increasing $b$ at sufficiently high $\Delta\gamma/\Bar{\gamma}$. Dotted lines in panel (b) represent the limits of $\mathcal{C}_2$ at high bias for the two extreme values of $\Delta\gamma$. These limits are based on the expressions of Sec. \ref{sec:analytic}. $\tau=\Bar{\gamma}=1$.}
    \label{fig:var}
\end{figure}

\begin{figure}[htpb]
    \centering
    \includegraphics[width=0.85\columnwidth, trim=18 10 18 0]{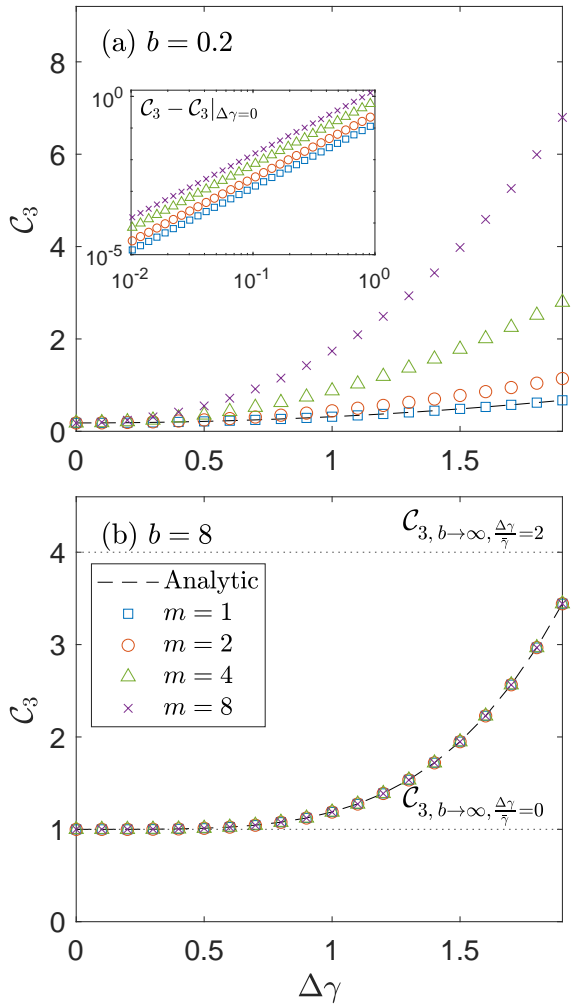}
    \caption{Scaled skewness as a function of $\Delta\gamma$ for varying segment lengths, $m_A=m_B\equiv m$. Dashed black curves represent the analytic result for $m_P=2$. Inset of panel (a): log-log plot of the skewness with its value at $\Delta\gamma=0$ subtracted off, showing that it scales quartically with $\Delta\gamma$ for varying $m$. Dotted lines in panel (b) represent the high-bias limits of $\mathcal{C}_3$ at the two extreme values of $\Delta\gamma$ for $m_A=m_B=1$. Parameter values are the same as in Fig. \ref{fig:var}.}
    \label{fig:skewness}
\end{figure}

\begin{figure*}[htbp]
    \centering
    \includegraphics[width=0.99\textwidth, trim = 95 15 95 20]{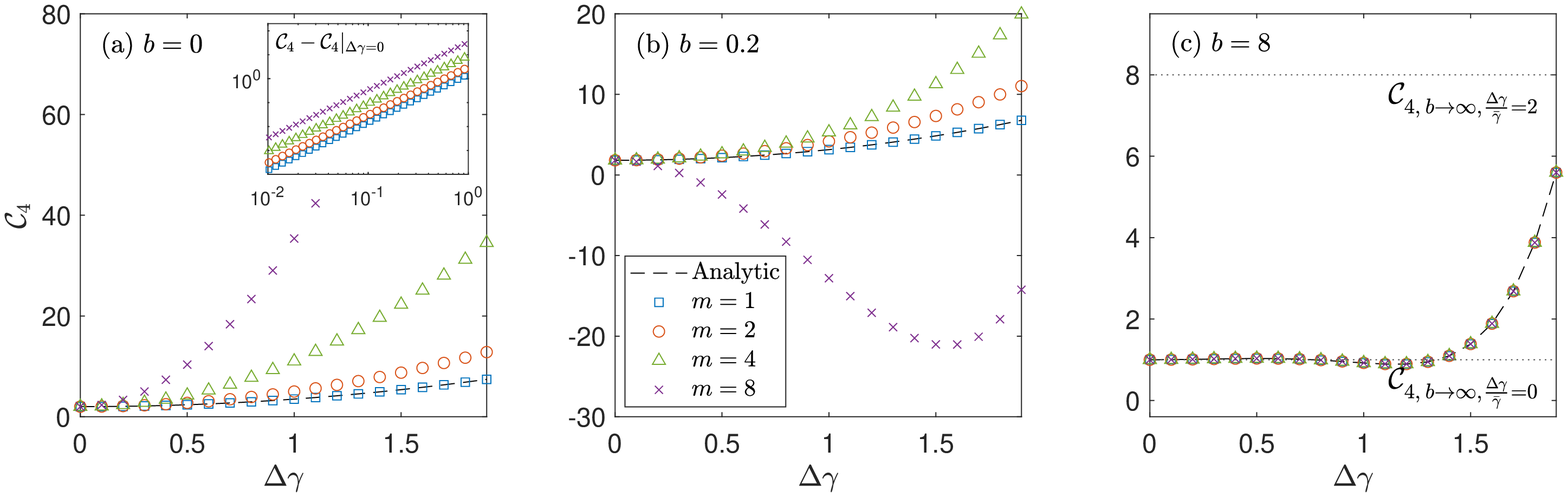}
    \caption{Numerically determined scaled kurtosis as a function of $\Delta\gamma$ for varying $m_A=m_B\equiv m$ at (a) zero, (b) low, and (c) high bias. The black dashed curve represents the analytic expression determined for the $m_A=m_B=1$ case. Inset of panel (a): log-log plot of the kurtosis with its value at $\Delta\gamma=0$ subtracted off, showing uniform scaling between different $m$ values. Dotted lines in panel (c) represent the high-bias limits of $\mathcal{C}_4$ at low and high $\Delta\gamma$, as determined for $m_P=2$ in Sec. \ref{sec:analytic}. Parameter values are the same as in Fig. \ref{fig:var}.}
    \label{fig:kurtosis}
\end{figure*}

\subsection{Simulations}
\label{sec:sim}

Simulations allow us to supplement our analytic results and go beyond the $m=1$ case. We now elaborate on these results. 
We calculate the scaled cumulants for chains with greater segment lengths by implementing full counting statistics numerically.

In agreement with Ref. \cite{Derrida}, simulations indicate that, for $m_A=m_B$, the expression derived analytically for the mean velocity, Eq.~(\ref{eq:mean}), holds independently of the segment length and $\Delta\gamma$. This is demonstrated in Fig. \ref{fig:mean}, where the markers indicating the numerically-obtained values of the first cumulant line up with the analytic curve for {\it all} segment lengths throughout the range of bias values, with nonzero $\Delta\gamma$.

Beyond the mean, higher order scaled cumulants do exhibit $\Delta\gamma$-dependence, even for the $m=1$ case, as demonstrated in Sec. \ref{sec:analytic}. Thus, we focus on plotting these quantities as a function of $\Delta\gamma$ to investigate how sensitively they express information about the structure of the chain, and how variation in the segment length impacts this behavior.

In general, simulations show that segment length has a substantial impact on the values and behavior of cumulants with $\Delta\gamma$, at low and intermediate bias. 
However, at high bias the cumulants no longer reflect a segment-size dependence, and they behave as they would in the case that segments were one-site long, but still {\it differently} from the analogous homogeneous random walk ($\Delta\gamma=0$ point on the graph).
%


In particular, this is demonstrated for the diffusion coefficient in Fig. \ref{fig:var}, where greater sensitivity to $\Delta\gamma$ is observed as the segment length increases, but only at low bias. Note that at zero bias (not depicted), simulations have shown that the diffusion coefficient goes to $2k^*$ for all $m_A=m_B$, as we showed analytically for $m_A=m_B=1$, exhibiting no $\Delta\gamma$-dependence at any segment length.

The inset of Fig. \ref{fig:var}(b) demonstrates the nonmonotonicity of $\mathcal{C}_2$ with increasing bias, discussed in Sec. \ref{sec:analytic}. As shown, sufficiently high $\Delta\gamma/\Bar{\gamma}$ leads the sign of $\partial\mathcal{C}_2/\partial b$ to change at a certain value of $b$. Interestingly, for the larger values of $m$ probed only in simulations, the behavior is the reverse of that for $m=1$, with $\mathcal{C}_2$ increasing at first, and then beginning to decrease with growing $b$ at sufficiently high $b$. All curves converge in the high-$b$ limit, as is consistent with the main plot.

The skewness, shown in Fig. \ref{fig:skewness}, behaves similarly to the diffusion coefficient, taking on values that grow even more rapidly with increasing $\Delta\gamma$. Once again, the zero-bias case is not depicted, as the skewness is an odd-order cumulant and always vanishes in this regime.

Finally, the $\Delta\gamma$-dependence of the kurtosis is demonstrated in Fig. \ref{fig:kurtosis}, including at zero bias. Away from the high bias limit, variations in the segment length lead to substantial variations in the value of the kurtosis, including a strong nonmonotonic behavior of the kurtosis with $\Delta\gamma$ at intermediate bias.

While the higher-order cumulants take on different values for different segment lengths at low bias, we have demonstrated additionally that the nature of the scaling is consistent between different values of $m$. For instance, as we showed in Sec. \ref{sec:analytic}, the diffusion coefficient for the $m=1$ case scales quadratically with $\Delta\gamma$. Our simulations show that this scaling is quadratic for larger $m$ as well, despite the form of the diffusion coefficient not matching exactly. This is demonstrated on the log-log plot in the inset of Fig. \ref{fig:var}(a). Analogous findings for the skewness and kurtosis are shown on Figs. \ref{fig:skewness} and \ref{fig:kurtosis}, respectively.


\section{Real-space simulations of the master equation} 
\label{sec:space}

Modular junctions under bias display rich, even nonmonotonic trends as a function of bias and $\Delta \gamma$, which we now aim to explain through
direct simulations of the
probability distribution functions.
We obtain the PDF of the random walk by numerically solving the master equation for site populations at different times. 
As an initial condition, we assume a probability of 1 to be at site $n=0$ and 0 elsewhere. We perform such simulations for long but finite chains ($\gtrsim$ 160 sites) with absorbing boundary conditions.
The simulation time is chosen long enough to observe a behavior corresponding to the steady state limit of the associated finite cycle. That is, we reach the situation of only the smallest-magnitude eigenvalue in the Liouvillian substantially contributing to the dynamics. From the other end, simulation time is limited to ensure that boundary effects do not come into play.

\begin{figure*}
    \centering
    \includegraphics[width=0.84\textwidth, trim=40 10 40 35]{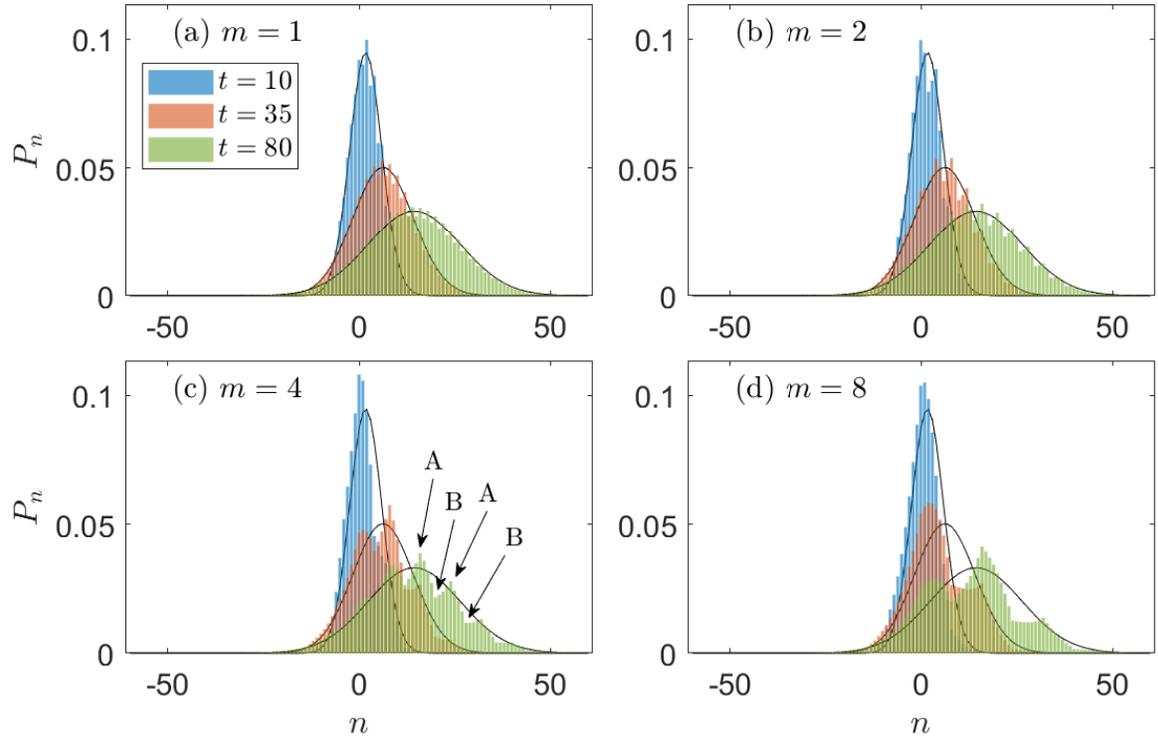}
\caption{Probability distribution functions at low bias 
obtained directly from the master equation for a 161-site chain. Each panel shows snapshots of the PDF at three different times for a given segment length, $n$. The black curves are the analogous probability distributions for the associated homogeneous random walk. $\Bar{\gamma}=\Delta\gamma=\tau=1$, $b=0.2$.}
    \label{fig:PDF_lowbias}
\end{figure*}

\begin{figure*}
    \centering
\includegraphics[width=0.84\textwidth, trim=45 10 45 10]{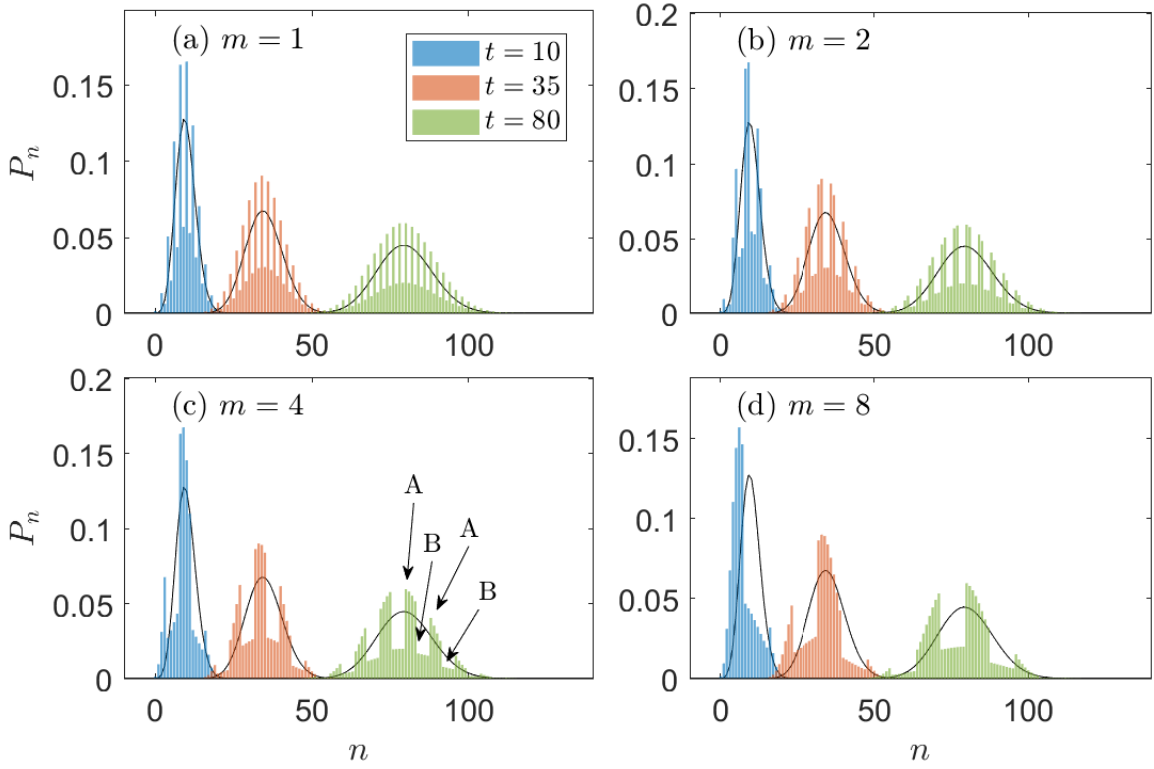}
\caption{Probability distribution functions at high bias 
obtained directly from the master equation for a 321-site chain (with most of the negative-$n$ region not shown). Each panel shows snapshots of the PDF at three different times for a given segment length, $n$. The black curves are the analogous probability distributions for the associated homogeneous random walk. $b=8$, representing the high bias regime. Parameter values are otherwise the same as in Fig. \ref{fig:PDF_lowbias}.}
    \label{fig:PDF_highbias}
\end{figure*}

We present the PDF for the random walk in Figs. \ref{fig:PDF_lowbias} and \ref{fig:PDF_highbias}
for the cases of low and high bias, respectively.
In each case, we study the PDF at different times, and for four different values of the segment size $m$: 1, 2, 4, and 8.

As shown in Fig. \ref{fig:PDF_lowbias}, at low bias, while the mean of the distribution remains in the same position as that for the associated homogeneous chain, there are additional features that grow more dramatic with increasing segment length $m$. These features amount to a series of local maxima and minima arising as a result of the modular structure, and are understood to account for the exotic behavior of the higher order cumulants.

In particular, population builds up in the left-most sites of the `A' segments, characterized by slower rates. 
Due these sites' positioning on the modular chain, population exits to the right at rate $k_A$, while it enters from the left at the faster rate $k_B$. In addition, the rate to exit to the left is suppressed by the factor $e^{-b}$. As such, population is generally fast to enter the A sites and slow to exit, accounting for the buildup of population at these sites observed in Fig. \ref{fig:PDF_lowbias}.
Conversely, at the left-most sites of `B' segments, we see a depletion of population due to the opposite effect. The transition rates into this state from A are relatively slow while the rate to exit to the left (towards A sites)  
is fast. A few examples of these population maxima and minima are labelled in Fig. \ref{fig:PDF_lowbias}(c) with the letters `A' and `B', respectively.
Overall, we observe a probability distribution that deviates from the smooth curve exhibited by the homogeneous random walk. This deviation is more dramatic with longer segment lengths.

The limit of high bias is exemplified in Fig. \ref{fig:PDF_highbias}, where we see similar buildup of population in the less rigid `A' segments and reduced population in the `B' segments due to the faster transitions out of these sites. The letters `A' and `B' label a few examples of this behavior in Fig. \ref{fig:PDF_highbias}(c). However, some of the very complex structure observed in the lower bias case is absent, due to the fact that reverse transitions are effectively eliminated.
A trajectory to reach site $n$ is understood as simply a sequence of $n$ steps forward, approximately half with waiting times characterized by the rate $k_A$ and the other half by $k_B$. The order at which these steps occur is determined by the value of $m$, but this no longer impacts the higher order cumulants of the distribution, explaining the indifference to $m$ that the diffusion coefficient, skewness, and kurtosis were shown to exhibit at high bias in Sec. \ref{sec:sim}. The probability distribution for two different segment lengths values line up at values of $n$ that are common multiples of the segment lengths.

In the Appendix, we calculate the cumulants from the real-space simulations of the PDF as a function of time. Finite time effects are rich \cite{Supriya1,Supriya2}; in the steady state we show that the scaled cumulants agree with results from Sec. \ref{sec:fcs}.

\section{Summary}
\label{sec:summ}
We have investigated the question of how the statistics of a random walk can be used to gain information about its underlying structure. Namely, we have examined random walks on one-dimensional modular chains, with repeating fast and slow segments, by determining how the cumulants of the population distributions over their sites, scaled by time, behave in comparison to those for analogous homogeneous random walks. We have found that the first cumulant, or mean velocity, always takes a form in the long time limit that matches the form for the homogeneous walk. Thus, measurements taking into account only the mean velocity at steady state are not sufficient to distinguish modular random walks from their homogeneous counterparts.

Studying the statistics in greater detail can, however, be an effective way to elucidate this very structure, with each of the higher-order cumulants discussed here reflecting the underlying modularity. This work probes cumulants as high as the kurtosis, which we found to be more 
expressive of modular structure than the diffusion coefficient and skewness; the kurtosis is nonzero and it differs in value from that of its homogeneous counterpart even in the zero-bias case. In the presence of bias, however, all cumulants beyond the mean velocity can be used to gain information about the underlying structure.

Population distributions themselves can also elucidate the impact that modular structure has on the behavior of random walks. They exhibit deviations from the Gaussian form expected for homogeneous random walks, with local maxima and minima occurring with the same periodicity along the chain as the variations in transition rates. These are due to local buildup and depletion of population due to the differing transition rates within one period of the chain's structure.

The present work considered classical transport---it is interesting to generalize these observations to quantum dissipative transport,
e.g., by using the formalism of quantum master equations with full counting statistic analysis \cite{esposito,Hava}. In this regard, it is intriguing to understand the role of quantum coherences in the behavior of noise going beyond
the homogeneous case
\cite{esposito05}
and beyond the second moment.
This task can be tackled by e.g., the unified quantum master equation,  \cite {Anton}, which was recently proved to be thermodynamically consistent in the steady state regime \cite{UnifedM}. 

Furthermore, the underlying structure of a random walk may not always be characterized by perfectly periodic variations in transition rates. Instead, one may consider a random walk on a disordered chain, with site-to-site transitions rates whose spatial variation is more random. The analogous problem has been studied for Brownian motion in continuous space, with disorder having notable impacts on the diffusion coefficient \cite{lindenberg11,lindenberg13}. Future work may investigate how this kind of underlying structure might be reflected in the higher order cumulants of the population distribution for random walks.

In addition, it would be interesting to study how higher order cumulants may reflect structure that is associated not just with transition rates, but also with the geometry of the network of states on which the random walk plays out. For instance, one may investigate how the presence of side chains or branches in the underlying network may be inferred through measurements of the statistics at steady state.




\begin{acknowledgments}
D.S. acknowledges the NSERC discovery grant and the Canada Research Chairs Program.
M.G. acknowledges support from the Ontario Graduate Scholarship and the NSERC Canada Graduate Scholarship-Doctoral. 
The authors acknowledge  Anton Zilman for fruitful discussions on kinetic networks. 
\end{acknowledgments}

\appendix

\section* {Appendix: Calculation of the cumulants from real-space simulations}
\renewcommand{\theequation}{A\arabic{equation}}
\renewcommand{\thesection}{A\arabic{section}}
\setcounter{equation}{0}  
\setcounter{section}{0}
\label{app:1}

\begin{figure}
    \centering
    \vspace{9mm}
    \includegraphics[width = 0.95\columnwidth, trim=40 10 40 30]{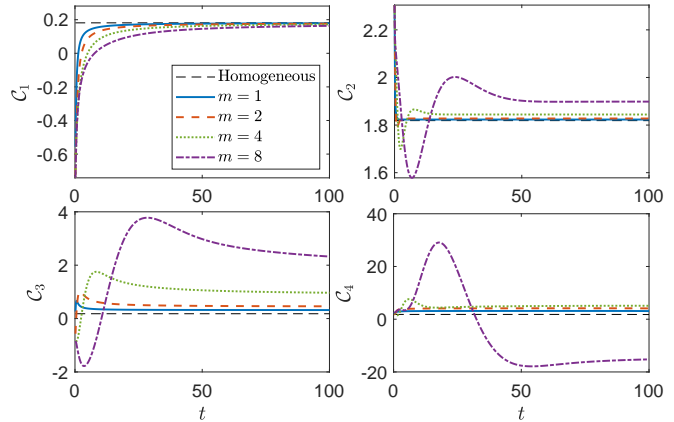}
    \caption{The first four scaled cumulants as a function of time, calculated directly from the solution of the master equation for a 161-site chain. Different curves represent different segment lengths $m$, with the black dashed lines representing the steady-state value of each cumulant for the associated homogeneous chain. $b=0.2$, representing the low-bias regime. $\tau=\Bar{\gamma}=\Delta\gamma = 1$.}
    \label{fig:cumulants_direct_lowbias}
\end{figure}

\begin{figure}
    \centering
    \includegraphics[width = 0.95\columnwidth, trim=40 10 40 10]{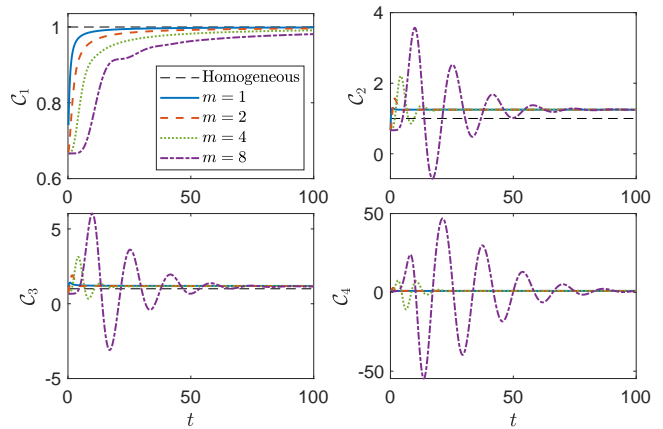}
    \caption{The first four scaled cumulants as a function of time, for various segment lengths, calculated directly from the solution of the master equation for a 321-site chain. The black dashed lines represent the steady-state value of each cumulant for the associated homogeneous chain. This reference line is hidden in the plot of $\mathcal{C}_4$ due to the large amplitude of its pre-steady-state oscillations setting the vertcal axis scale. $b=8$, representing the high-bias regime. Parameter values are otherwise the same as in Fig. \ref{fig:cumulants_direct_lowbias}.}
    \label{fig:cumulants_direct_highbias}
\end{figure}
%
We use the numerical solutions of Sec. \ref{sec:space} to calculate the scaled cumulants as a function of time. In particular, we calculate the moments of the distribution over $n(t)$, which are represented by single angle brackets,
\be 
    \la n^k(t)\ra = \sum_n n^k P_n(t).
\ee
The summation is done over the many sites included in the simulation. We then derive the scaled cumulants via Eq.~(\ref{eq:cumulants_def}), which amounts to the following expressions when written in terms of the moments:
\begin{align}\label{eq:cumulants_direct}
    \mathcal{C}_1(t) &= \frac{\la n(t) \ra}{t}
    \nonumber\\
    \mathcal{C}_2(t) &= \frac{\la (n(t) - \la n(t)\ra)^2 \ra}{t}
    \nonumber\\
    \mathcal{C}_3(t) &= \frac{\la (n(t) - \la n(t)\ra)^3 \ra}{t}
    \nonumber\\
    \mathcal{C}_4(t) &= \frac{\la (n(t) - \la n(t)\ra)^4 \ra - 3\la (n(t) - \la n(t)\ra)^2\ra^2}{t}.
\end{align}
Note that the scaled cumulants here are time-dependent quantities that depend on the choice of initial state. In the long time limit, they converge to the steady state values calculated by the method outlined in Sec. \ref{sec:fcs}.

The scaled cumulants as calculated in this manner are shown in Figs. \ref{fig:cumulants_direct_lowbias} and \ref{fig:cumulants_direct_highbias} for low bias ($b=0.2$) and high bias ($b=8$), respectively. The 
parameter values match those of Figs. \ref{fig:PDF_lowbias} and \ref{fig:PDF_highbias} exactly. 
The probability distributions $P_n$ for the 161- and 321-site chains do not reach steady state within the time frame of the simulation.  
%
However, these systems can be observed to reach a quasi-steady state as each of the scaled cumulants, $\mathcal{C}_k$, approach their asymptotic value. The timescale for this to occur is the timescale for the analogous bipartite finite chain to reach its steady state, given by the largest nonzero eigenvalue of the rate matrix for this system, as discussed in Secs. \ref{sec:MM} and \ref{sec:fcs}. This timescale is observed to grow substantially with the segment length $m$--in the context of the finite cycle, this is intuitive as a greater number of steps are needed for the population to spread out over all the states.

The asymptotic values of the scaled cumulants line up with the steady-state values shown in Figs. \ref{fig:var}-\ref{fig:kurtosis} at $\Delta\gamma=1$, with visible deviations from the value predicted for the homogeneous chain. Before reaching steady state, however, these quantities vary quite dramatically, particularly for larger $m$.

\end{document}